\documentstyle[12pt,aasms4,flushrt]{article}

\def\deg{^{\circ}}
\def\degsq{~{\rm deg}^2}
\def\minsq{~{\rm arcmin}^2}

\def\itb{{\it B\/}}
\def\iti{{\it I\/}}
\def\itk{{\it K\/}}
\def\itu{{\it U\/}}
\def\kprime{$K^{\prime}$}
\def\bband{{\it B\/}}
\def\iband{{\it I\/}}
\def\kband{{\it K\/}}

\def\bmi{{($B-I$\/)}}
\def\imk{{($I-K$\/)}}

\def\ang{~{\rm \AA}}
\def\msun{~\rm M_{\odot}}
\def\lsun{~\rm L_{\odot}}
\def\kms{~{\rm km\ sec}^{-1}}

\def\lal{{\rm Ly}\alpha}
\def\halpha{{\rm H}\alpha}
\def\lstar{\ifmmode {L_{\ast}}
        \else {$L_{\ast}$}\fi}
\def\mk{M_{\rm K}}
\def\mkstar{M_{\rm K\ast}}
\def\mbstar{M_{\rm B\ast}}
\def\phistar{\phi_{\ast}}
\def\fnu{\phi_{\nu}}
\def\kstar{K_{\ast}}
\def\bstar{B_{\ast}}

\def\hnought{{\rm H}_0}
\def\qnought{\ifmmode q_0
    \else $q_0$\fi}
\def\qno{\ifmmode q_0
    \else $q_0$\fi}

\def\ten#1{\ifmmode 10^{#1}
    \else $10^{#1}$\fi}
\newcount\hours
 \newcount\minutes
  \def\SetTime{\hours=\time
         \global\divide\hours by 60
         \minutes=\hours
         \multiply\minutes by 60
         \advance\minutes by-\time
         \global\multiply\minutes by-1 }
 \SetTime
 \def\now{\number\hours:\ifnum\minutes<10 0\fi\number\minutes}

\begin{document}

\title{New Insight on Galaxy Formation and Evolution from Keck Spectroscopy
  of the Hawaii Deep Fields}
\author{Lennox L. Cowie,\altaffilmark{1,2} Antoinette Songaila,\altaffilmark{1} 
  and Esther M. Hu\altaffilmark{1,2}}
\affil{Institute for Astronomy, University of Hawaii, 2680 Woodlawn Dr.,
  Honolulu, HI 96822\\
  cowie@ifa.hawaii.edu, acowie@ifa.hawaii.edu, hu@ifa.hawaii.edu}
\and
\author{J. G. Cohen}
\affil{Palomar Observatory, Mail Stop 105-24, California Institute of 
  Technology, Pasadena, CA 91125\\
  jlc@astro.caltech.edu}
\altaffiltext{1}{Visiting Astronomer, W. M. Keck Observatory, jointly
  operated by the California Institute of Technology and the University of
  California.}
\altaffiltext{2}{Visiting Astronomer, Canada-France-Hawaii Telescope,
  operated by the National Research Council of Canada, the Centre National
  de la Recherche Scientifique of France, and the University of Hawaii.}

\slugcomment{Submitted to {\it Astronomical Journal\/}, January 4, 1996\ \ 
Second Revision May 21, 1996}

\begin{abstract}
We present the results of spectroscopic studies with the LRIS spectrograph
on Keck of two of the Hawaii deep survey fields.  The 393 objects observed
cover a 26.2 arcmin$^2$ area and constitute a nearly complete sample down
to $K=20$, $I=23$, and $B=24.5$.  The rest-frame \itk-band luminosity
function and its evolution with redshift are described.  Comparisons are
made with other optically selected (\itb\ and \iti) samples in the
literature, and the corresponding rest-frame \itb-band luminosity function
evolution is presented.  The \itb-band counts at $B\sim24$ are shown to be
a mixture of normal galaxies at modest redshifts and galaxies undergoing
rapid star formation, which have a wide range of masses and which are
spread over the redshift interval from $z=0.2$ to beyond $z=1.7$.

The luminosity functions, number counts, and color distributions at optical
and IR wavelengths are discussed in terms of a consistent picture of the
star-forming history of the galaxy sample.  [\ion{O}{2}] emission-line
diagnostics or rest-frame ultra-violet--infrared color information are used in
combination with rest-frame absolute \itk\ magnitudes to construct a
``fundamental plane'' in which the evolution of the global star-formation
rate with redshift can be shown, and we find that the maximum rest-frame
\itk\ luminosity of galaxies undergoing rapid star formation has been
declining smoothly with decreasing redshift from a value near \lstar\ at
$z>1$.  This smooth decrease in the characteristic luminosity of galaxies
dominated by star formation can simultaneously account for the high
\itb-band galaxy counts at faint magnitudes and the redshift distribution
at $z<1$ in both the \itb- and \itk-selected samples.  Finally, the overall
\itk-band light density evolution is discussed as a tracer of the baryonic
mass in stars and compared with the overall rates of star formation
inferred from the rest-frame ultraviolet light density as a function of
redshift.

\end{abstract}

\section{Introduction}

In order to understand the formation and evolution of galaxies we require
very large complete redshift samples to faint magnitudes.  Such large
samples have recently been obtained in the optical with both blue and red
selections ($B < 24$: \markcite{gla95a}Glazebrook et al.\ 1995a,
\markcite{ell96}Ellis et al.\ 1996; $I < 22$: the CFRS Survey -- see, e.g.,
\markcite{lil95a}Lilly et al.\ 1995a).  These samples contain several
hundred to over one thousand galaxies.  However, optically selected samples
are severely biased against earlier galaxy types at the higher redshifts
and so are rather poorly suited to studying the evolution beyond about $z =
1$\ where, even in the \iti\ band, the K corrections become very large and
many of the galaxy types disappear from the selection.

Infrared-selected samples minimize this problem because the K corrections 
are relatively invariant to galaxy type and remain small (indeed
negative) to high redshift.  They are therefore well suited to
studying high-redshift populations and are our best method of determining
the formation process.  Two quite large faint \itk\ (2.2~micron) or
\kprime\ (2.1~micron) samples have recently been published, by Glazebrook
et al.\markcite{gla95b} (1995b) and Songaila et al.\markcite{son94} (1994).  
However, the former sample extends only to $K = 17$, not deep enough to
reach the $z \gtrsim 1$\ population, and the latter sample, which reaches
$K = 20$, is somewhat incomplete at the faint end and covers a quite small
area at $K \gtrsim 18$.

In this paper we shall describe a much larger, and very faint infrared
sample (area = $26.2\minsq$, $K \le 20$) of 254 objects which we have
observed using the LRIS spectrograph (\markcite{oke95}Oke et al.\ 1995) 
on the Keck 10m telescope.  The sample consists of all objects in two
$6\arcmin \times 2\arcmin$\ strips, and is nearly complete to $K = 19.5$.
To complement the \itk-selected sample we have also observed the $I <
23$\ or $I < 22.5$\ and $B < 24.5$\ objects in the two fields, bringing
the total to 393 objects in all.

These three color-selected redshift samples are extremely powerful tools for
understanding the evolution of galaxies.  We first describe the evolution of
the luminosity functions and light densities in the various rest-frame colors
(\itb\ through \itk) out to a redshift of 1.6 and compare the data to the
optical and brighter infrared samples in the literature.  Next, we use these
results to study the evolution of galaxies in a parameter space characterized 
by the absolute rest \itk\ magnitude, $\mk$, in combination with either the 
[\ion{O}{2}] or $\halpha$\
equivalent widths or the rest-frame ultraviolet--infrared color, $UK(AB)$, and
the redshift.  This is equivalent to a parameterization of each galaxy as a
function of mass (shown by $\mk$) and star formation rate, and we find that
the galaxy population shows a rapid migration in this plane as a function of
$z$.  At $z > 1$, galaxies with near-$\lstar$\ infrared luminosities have
[\ion{O}{2}] equivalent widths or $UK(AB)$\ colors that correspond to very
rapid star formation.  At lower redshifts, the maximum rest \itk\ luminosity
of such ``forming'' galaxies drops smoothly until at the present epoch there
are only a few, generally very low luminosity, ``forming'' galaxies.

We shall refer to this remarkably smooth downward evolution in the maximum
luminosity of rapidly star-forming galaxies as ``downsizing,'' since the
assembly of the upper end of the galaxy luminosity function is occurring
from the top down with decreasing redshift; it is this phenomenon that
accounts for the simultaneous rapid evolution in the colors of galaxies and
in the normalization of the luminosity function (or equivalently in the
\itb- and \itk-band galaxy counts) while leaving the redshift distribution
at $z < 1$\ well characterized by a no-evolution model (\markcite{bro88}%
Broadhurst et al.\ 1988, \markcite{col90}Colless et al.\ 1990, \markcite{ste94}%
Steidel et al.\ 1994).  This solution, which relies on the \lstar\
galaxies being in place (though still undergoing some star formation at a
declining rate) at $z < 1$ while sub-$\lstar$\ galaxies are still forming, 
was originally suggested in \markcite{bro88}Broadhurst et al.\ (1988) to
deal with the $z \sim 0.3$\ blue galaxies; however, we now see that the
faint ($B \gtrsim 24$) blue counts contain a variety of galaxies
``forming'' over a range of redshifts, including massive galaxies at $z >
1$.  This range of blue galaxies in redshift and luminosity fully accounts
for the faint blue excess in the number counts.

Because the interpretation draws on all the information in all three
colors, we have chosen not to divide the paper, which is consequently
somewhat long.  Many readers may wish to skip the technical details of the
observations given in Sec.\ \ref{sec:observ}.  The reader who is primarily
interested in the interpretation may further skip to Sec.\
\ref{sec:interpret}, which is intended to be self-contained.

\section{Observations}\label{sec:observ}

The sample was defined as all objects with $K < 20$, $I < 22.5$
(Kron-Cousins), or $B < 24.5$ in two approximately $6\arcmin\times
2\arcmin$ areas (exact combined area = $26.2\minsq$) surrounding the Hawaii
deep survey fields SSA~13 and SSA~22 (\markcite{cow94}Cowie et al.\ 1994, 
\markcite{son94}Songaila et al.\ 1994) and lying along the E-W direction.  
The SSA~22 sample was extended to include all $I = 23$\ objects.  The 1950
coordinates for each object (absolute accuracy better than $0\farcs4$) are
given in Tables \ref{tbl-1a} and \ref{tbl-1b}, together with their \kband,
\iband, and \bband\ magnitudes and their corresponding numbered
identifications on \iband\ finding charts of the field, shown in
Figs.\ \ref{fig:1a} and \ref{fig:1b}.  The \kband, \iband, and
\bband\ magnitudes are $3\arcsec$\ diameter aperture magnitudes corrected
to $6\arcsec$\ diameter with an average offset.  The object selection
process is described in the previous papers.  Tables \ref{tbl-1a} and
\ref{tbl-1b} are ordered by R. A.\ and for each field list first the $K<
20$\ sample, then the additional objects which complete the
\iband\ sample, and finally those added to complete the \bband\ sample are
given.  A small fraction of the objects are not detected in one of the
colors at the limiting $1~\sigma$\ magnitudes of $K = 21.8$, $I = 25.3$,
and $B = 26.3$.  A negative aperture flux is indicated by a magnitude
corresponding to the absolute flux preceded by a negative sign.  The
SSA~13 catalog contains 174 objects and the SSA~22 catalog 219 objects;
the larger number of objects in the SSA~22 field is partly a result of
the fainter \iband\ selection and partly of the lower galactic latitude
of SSA~22 ($-44\deg$) compared with SSA~13 ($+74\deg$), which results in
a larger fraction of stars.

Existing spectroscopy of a number of galaxies in these fields is summarized
in \markcite{son94}Songaila et al.\ (1994).  Most of the remaining objects 
without existing spectra were observed with the LRIS multi-object spectrograph 
(\markcite{oke95}Oke et al.\ 1995) on the Keck 10 meter telescope in a
number of runs from 1994 August to 1995 September.  Each mask on the
spectrograph contained from 22 to 26 $1\farcs4\times 12\arcsec$\ slits in
the field area.  The 300 $\ell$/mm grating used gives roughly $5000\ang$\ of
wavelength coverage from $\sim 5000-10,000\ang$\ with a pixel size of
$0\farcs21$ along the slit and $2.5\ang$ in the wavelength direction.  The
blue wavelength cutoff for the multiple slits ranged in practice from
$4300\ang$\ to $5700\ang$, depending on the position of the slit on the
mask.  With the wide $1\farcs4$\ slit, the resolution was $16\ang$.

Because the CCD chip is extremely flat, we were able to adopt a relatively
simple but robust observing procedure (a variant on self-flattening) which
allowed us to avoid the extremely difficult and time-consuming problems of
flat-fielding, which is a considerable advantage because of internal
flexure problems and the very high overheads associated with spectrograph
setup, motions, and CCD readouts.  Each mask was observed with objects
positioned at the center of the slit and then at $\pm 2\farcs5$\ along the
slit.  Each such exposure was 20 minutes in duration giving a total
integration on each mask of 1 hour.  The minimum of the three frames at
each pixel was used to perform a first-pass sky subtraction and then the
frames were registered and median added to form the spectral image.  A
median $3\times 3$\ spatial filter was then used to identify and eliminate
the very small number of cosmic rays present in the median summed frame.
Next, the geometric distortion was removed from the spectral images using
fits to the edge positions of the dispersed slits, as a function of
wavelength.  Third order polynomial fits to distortion-corrected
observations of an Ar lamp taken at the beginning and end of the night were
used to determine the wavelength scales, but, to allow for flexure, were
offset to match the night-sky spectrum in each slit.  A final linear sky
subtraction in each slit completed the reduction.  
The night sky subtraction is generally excellent even at the
redder wavelengths.

The final one-dimensional extraction was made using a summation weighted by
the slit profile (optimal extraction) and was approximately
flux calibrated to $f_{\nu}$\ using observations of white dwarf standards
taken in the same configuration.  Redshifts were then measured for each of
the objects by two of the authors independently.  Where a redshift was
considered secure (i.e. definitely identified by both observers) it was
entered into the catalog of Tables \ref{tbl-1a} and \ref{tbl-1b} and
generally excluded from future mask design unless there were no further
unidentified objects at this position on the E-W slit axis of the mask.
However, dubious or unidentified objects were included in additional masks
until such time as the summed spectrum produced a reliable identification.
The longest exposure on a single object was six hours.  Objects which still
do not have a robust identification are marked by zeros in the catalog of
Tables \ref{tbl-1a} and \ref{tbl-1b}, while stars are marked with $-2$\ and
unobserved objects with $-1$.  For the completeness analysis, we consider
unidentified objects and unobserved objects to be equivalent, but for some
statistical tests it may be appropriate to distinguish them.

Because of the large number of objects, it is not practicable to show all
the spectra.  Instead we show in Figs.\ \ref{fig:2a} and \ref{fig:2b}
every 10th object, excluding stars and unobserved objects (designated
$-2$\ and $-1$\ in the catalog).  The spectra are presented in the rest
frame with the shaded regions showing the positions of strong sky features
(the atmospheric bands and the stronger night sky lines).  The redshift is
shown at the top of the plot, the number of the object in Tables
\ref{tbl-1a} or \ref{tbl-1b} and Figs.\ \ref{fig:1a} or \ref{fig:1b} at
the lower left, and the \kband\ and \bband\ magnitudes at the lower right.  

Finally, we show in Fig.~\ref{fig:3} the completeness as a function of
magnitude in each color, as well as the number of objects in the catalog.
For the SSA~13 field we have included objects only to $B < 24$\ because a
relatively large number of objects with $B = 24-24.5$\ have yet to be
observed in this field.  The samples are reasonably complete to $K = 19.5$,
$I = 22.5$, and $B = 24.5$.

\section{The $K$-Band Sample}

\subsection{$K-z$ Relation}

We have combined the present data for $K = 18-20$\ with the much larger
area samples to $K < 18$\ given in \markcite{son94}Songaila et al.\ (1994).  
The Songaila et al.\ sample ranges in area from $1.54\degsq$\ at $K <
14.5$\ to $89\minsq$\ at $K < 18$\ and is nearly fully complete at these
magnitudes.  The combined sample contains 416 objects, of which 31 are
stars and 346 are galaxies with well determined redshifts.  From
Fig.~\ref{fig:3} it can be seen that the sample is 77\% 
complete at $K = 19-19.5$, but only 57\% complete at $K = 19.5-20$.

We show the redshift-magnitude relation in Fig.~\ref{fig:4}, with filled
triangles giving the Songaila et al.\ sample and filled squares showing data
from the present catalog.  The diagram has been extended to $K = 21$, but
beyond $K = 20$\ (the vertical dashed line) there is, of course, a strong
color bias in the sample since it contains only objects with $I < 23$\ or $B <
24.5$.  Unidentified objects are not shown in this plot.  Other available
samples are also shown in Fig.~\ref{fig:4} --- the open symbols show the
\markcite{mob86}Mobasher et al.\ (1986) and \markcite{gla95b}Glazebrook et
al.\ (1995b) samples, and we show unidentified objects in the present sample
as crosses at a nominal redshift.  The methods for determining the magnitudes
are somewhat different in each of these samples and this requires some
consideration.  The various Hawaii samples use apertures of sufficiently large
angular diameter so that small average corrections may be applied to obtain
approximate total magnitudes, whereas \markcite{mob86}Mobasher et al.\ (1986)
used corrected isophotal magnitudes.  For comparison purposes we have
therefore used the \markcite{gla95b}Glazebrook et al.\ (1995b) metric
magnitudes measured in a $40\,h_{50}^{-1}~{\rm kpc}$\ aperture (where
$h_{50}=\hnought/50\kms$ Mpc$^{-1}$), which are closer to the total magnitudes
used here and by Mobasher et al., and we have cut off their sample at $K =
17$.  Because of the different methodologies, there may be internal
dispersions as large as 0.2 to 0.3 mags (e.g., \markcite{cow94}%
Cowie et al.\ 1994, hereafter Paper~I), and smaller systematic offsets
between the samples.

Fig.~\ref{fig:4} can be used to see many of the conclusions of this section
in a semi-quantitative fashion.  Here we have plotted two curves, the lower
one for a $0.1\lstar$\ galaxy and the upper one for a $3\lstar$\ galaxy,
for $\mkstar = -25.0$\ ($+5 \log_{10}\,h_{50}$) where we have computed
$K$\ versus $z$\ using $\qno = 0.5$\ and the K-correction for an Sb galaxy
given in Paper~I and overlaid these on the $K-z$ distribution of objects
from both the present sample and the Glazebrook et al.\ and Mobasher et
al.\ samples.  These extrema should roughly bound the distribution of
redshifts, given an $\alpha = -1$\ Schechter function, as indeed they do at
lower redshifts.  However, at fainter magnitudes the redshifts begin to
fall systematically low.  This may be more clearly seen in Fig.~\ref{fig:5}, 
where we show the median redshift versus $K$\ magnitude for the present
sample and those of Glazebrook et al.\ and Mobasher et al.  The
$1\sigma$\ error bars are calculated in a highly conservative fashion by
using the median sign method, allowing for the possibility that
unidentified or unobserved objects may be at either high or low redshift.
At faint magnitudes the data points lie between the predictions of a no
luminosity evolution model computed here for a Schechter function with
$\mkstar = -25.0 + 5\log_{10}\,h_{50}$, $\alpha = -1$, and $\qno =
0.5$\ and $\qno = 0.02$\ (solid and dashed lines, respectively) and the
lower values predicted by a strong merger model such as those of
Rocca-Volmerange \& Guiderdoni\markcite{roc90} (1990), Broadhurst et
al.\markcite{bro92} (1992), and Carlberg\markcite{car92} (1992) (whose
model's predictions are shown as a dotted line on the figure), but are well
below even a mild passive luminosity evolution model, illustrated here by
the dash-dot line.  The $K$\ luminosity per galaxy would therefore appear
to be invariant, or even slightly falling, at higher redshifts.

\subsection{The evolution of the $K$-band luminosity function}

For each galaxy we have determined a rest-frame absolute \kband\ magnitude
using the K-correction for an Sb galaxy given in Paper~I as
\begin{equation}
   M_{\rm K} = m_{\rm K} - 5\log\left ( {{d_{\rm L}} \over {10~{\rm pc}}}
   \right ) - {\rm K}(z) \label{eq:1}
\end{equation}
where $d_L$\ is the luminosity distance and $m_{\rm K}$\ the apparent
\kband\ magnitude.  Between $z = 0$\ and 1, the use of a type-independent
K$(z)$\ correction is justified for the \kband\ band and introduces at most
a 0.1 mag error for an elliptical and 0.2 mag for an Im, errors smaller
than or comparable to the photometric uncertainties of $\sim 0.2$\ mag in
the catalogs.  This type invariance of the K correction greatly simplifies
the analysis.

The sample was next sorted by absolute magnitude from the most negative
(most luminous) to the most positive values, and the observable comoving
volumes between redshifts $z_1$\ and $z_2$\ computed for each galaxy as
\begin{equation}
   V_j = {c \over \hnought} \sum_i A_i \int^{z_{ui}}_{z_{li}}
   {{d_l^2\,dz} \over {(1+z)^3 (1+2q_0z)^{0.5}}} \label{eq:2}
\end{equation}
where $z_{li}$\ is the maximum of $z_1$\ and the minimum redshift at
which object $j$ lies in the $i$th magnitude range, and $z_{ui}$\ is
the minimum of $z_2$\ and the maximum redshift at which object $j$\
lies in the $i$th magnitude range.  $A_i$ is the observed area in the
$i$th magnitude range.  We then constructed the cumulative \kband-band
luminosity per unit comoving volume to an absolute magnitude $\mk$\
corresponding to object $k$\ as
\begin{equation}
   \lambda_{\rm K} \equiv \lambda (\mk ) = 
                       \sum_{j=1}^k {{L_j} \over{V_j}}\quad , \label{eq:3}
\end{equation}
which we shall refer to as the cumulative luminosity density, and
the cumulative number per comoving volume
\begin{equation}
   n(\mk ) = \sum_{j=1}^k {{1} \over{V_j}} \quad .\label{eq:4}
\end{equation}
Here $L_j$\ is the \kband-band luminosity of object $j$\ in solar
units, with $M_{\rm K \odot} = 3.41$.

This cumulative expression allows us to display the form of the luminosity
function without binning or parametric fits, and the smoothness of the
function may be used to visually estimate the degree of uncertainty.  The
principal demerit of this approach is that the errors are no longer
independent.  

In Fig.~\ref{fig:6a} we plot the cumulative rest-frame \itk-band luminosity
density versus $\mk$\ computed for the redshift ranges [0,0.3] (crosses),
[0.3,1] (boxes), and [1,1.6] (diamonds) for $\qno = 0.02$\ and 0.5; while
in Fig.~\ref{fig:6b} we show the corresponding number of objects in each
magnitude bin.  The
median redshift is 0.18 in the lowest redshift interval, 0.51 in the $z$ =
[0.3,1] interval, and 1.10 in the $z$ = [1,1.6] range.  The most striking
point is that the faint-end asymptotic luminosity density has risen by a
factor of 1.4 to 1.6 (depending on the assumed \qno) between the two lower
redshift ranges, a point to which we shall return in the next section.  Here
we concentrate on the shape of the luminosity function and, in order to
visually compare the shapes of the distributions in the two redshift
intervals, we have renormalized the $z$ = [0.3,1] distribution to have the
same light density at $\mk = -21 + 5 \log_{10} h_{50}$\ (shown by the solid
line).  It can then be seen that the solid line is shifted to slightly
fainter magnitudes for $\qno = 0.5$\ and is invariant for $\qno = 0.02$.
Because the highest redshift bin extends only over a small absolute
magnitude interval, it is not possible to compare the shape in this way, but
it is also consistent with an invariant or declining $\mkstar$.  In
particular, galaxies with $\mk < -26 + 5 \log_{10}h_{50}$\ have $K <
18.5$\ at $z = 1.6$\ and, therefore, lie within the essentially completely
identified sample; it must be, then, that the deficiency of luminous
objects at these high redshifts is not a selection effect.

We have formally fitted a Schechter function to the cumulative number
distribution in the two lower redshift bins using the Kolmogorov-Smirnov
test to determine 68\% confidence contours.  These are shown in
Fig.~\ref{fig:7} for the two parameters, $\alpha$\ and $\mkstar$.  The best
fit to the low-redshift interval gives $\alpha = -1.2$\ and $\mkstar = -
25.0 + 5 \log_{10}h_{50}$.  At higher redshift for $\qno = 0.5$\ the best
fit is $\alpha = -1.3$, $\mkstar = -25.0 + 5 \log_{10}h_{50}$, while for
$\qno = 0.02$\ it is $\alpha = -1.4$, $\mkstar = -25.4 + 5
\log_{10}h_{50}$.  Given the statistical and photometric uncertainties, the
results are best described as indicating an invariant, or, depending on the
assumed $\qno$, a very slightly decreasing, galaxy luminosity.

The present results agree well with those of Mobasher et al.\markcite{mob93}
(1993) who find $\alpha = -1$\ and $\mkstar = -25.0 \pm 0.3 + 5
\log_{10}h_{50}$\ in their low-redshift bright sample.  However, Glazebrook
et al.\markcite{gla95} (1995b) find $\mkstar = - 24.25 \pm 0.13 + 5
\log_{10}h_{50}$\ at $z = 0.0 - 0.4$\ and $\mkstar = -24.91 \pm 0 .15 + 5
\log_{10}h_{50}$\ at $z = 0.4 - 0.8$, where they force fit in both cases to
an $\alpha = -1$\ power law.  At the higher redshift this agrees well with
the present analysis but the $\mkstar$\ is considerably fainter at low
redshift than the value determined here even when allowance is made for
possible differences ($\sim 0.3$\ mag) between the $40\, h_{50}^{-1}\ {\rm
kpc}$\ metric magnitudes used by Glazebrook et al.\ and the present
magnitudes.  This led Glazebrook et al.\ to conclude that there was positive
luminosity evolution with increasing $z$, in disagreement with our result.
As Glazebrook et al.\ note, this difference may result from the fluctuations
inherent in small samples; the low number of super-$\lstar$\ galaxies can
result in biases in $\mkstar$.  Our $z = [0,0.3]$ bin contains 138 galaxies,
the Mobasher et al.\ sample 95 galaxies, and the Glazebrook et al.\ $z <
0.4$\ bin 98 galaxies.

In order to understand the disagreement better, we have therefore
performed the same analysis on the combined data sets.  With 296 galaxies
in the lower redshift $z = [0,0.3]$ bin, we find a best fit of $\mkstar =
-24.8 + 5 \log_{10}h_{50}$\ and $\alpha = - 1.1$.  In the higher redshift
bin with 232 galaxies, we find $\mkstar = - 24.9 + 5 \log_{10}h_{50}$\ and
$\alpha = -1.3$\ for $\qno = 0.5$\ and $\mkstar = -25.1 + 5
\log_{10}h_{50}$, $\alpha = -1.3$\ for $\qno = 0.02$.  The results are
again consistent with shape invariance ($\qno = 0.02$) or a very slight
dimming with increasing redshift ($\qno = 0.5$).

An alternative presentation is given in Fig.~\ref{fig:8}, where we have
directly computed the comoving luminosity functions in four redshift
intervals using the $1/V$\ method for $\qno = 0.5$.  Poisson errors are
shown, corresponding to the number of objects in each magnitude bin.  Where
no objects were detected in a bin, the $1~\sigma$\ upper limit is shown with
a downward pointing arrow.  Once again we have shown the Hawaii data only
(solid squares) and the combined data (open squares).  The fits (solid line
for the Hawaii data, dashed line for all data) are shown for a $\mkstar =
-25.0 + 5 \log_{10}h_{50}$, $\alpha = -1.25$\ Schechter function, which
provides an acceptable fit in all four redshift intervals (irrespective of
which data set is used) with a suitable choice of the normalization,
$\phi_{\ast}$.  The dotted line shows a reference value of $\phistar =
10^{-3}\,h_{50}^3~{\rm Mpc}^{-3}$.  The normalization lies at or below this
value in the $z < 0.1$\ and $z = 0.1 - 0.2$\ bins, rises to roughly a factor
of two higher in the $z = 0.2-0.6$ bin, and then begins to drop again at $z
= 0.6-1.0$.

\subsection{The evolution of colors and emission line strengths in the \itk\ 
sample}\label{subsec:col-line-evol}

The \imk\ versus \itk\ diagram for the present sample (\itk- or
\iti-selected) is shown in Fig.~\ref{fig:9a} with objects marked as stars
(asterisks), identified galaxies at $z < 1$ (filled boxes), galaxies at $z
> 1$\ (boxes surrounded by open boxes), and unidentified objects (pluses).
The error bars are $\pm 1\sigma$\ errors in \iti\ only.  The figure
demonstrates that the incompleteness in the \itk\ sample is primarily in
objects with redder \imk\ colors.  However, as has been previously
emphasized in Cowie et al.\ (\markcite{cow94}1994, \markcite{cow95b}1995b), 
blue objects dominate the \itk\ counts and many $z > 1$\ galaxies are
contained in the bluer objects near $K = 20$.  There are also relatively few
objects (less than eleven in this sample) which can have colors redder than
an Sb galaxy at $z > 1.5$\ ($I - K \ge 4.5$) even though an $\mkstar =
-25.0$\ Sb galaxy would have $K = 19.5$\ at $z = 1.5$.

This point is further emphasized in Fig.~\ref{fig:9b} where we show the
($I- K$) colors of a $K < 22.5$\ sample centered on the deep HST \iti-band
images of Cowie et al.\markcite{cow95b} (1995b) which lie at the centers of
the spectroscopic strips.  In these areas very much deeper \itk\ band images
($1~\sigma\ K = 24.5$\ in SSA~22 and $1~\sigma\ K = 23.5$\ in SSA~13) have
been obtained with the QUIRC $1024 \times 1024$\ IR camera at CFHT.  We show
all objects in a $3\minsq$\ area with $K < 22.5$\ in SSA~22 and with $K <
21.5$\ in a $3\minsq$\ area in SSA~13.  Even in this extremely deep sample
there are very few red objects (six with $(I - K) > 4.5$) and the great bulk
of \itk-selected objects are extremely blue in $(I - K)$.

To investigate the origin of this blueing trend within the present
spectroscopic data, we can look at either the emission-line characteristics
of the galaxies or their rest-frame ultraviolet colors.  For each galaxy we
measured the rest-frame equivalent width of the [\ion{O}{2}] $3727\ang$\
emission line (EW([\ion{O}{2}])) or, in cases where the redshifted
[\ion{O}{2}] lay blueward of our observed spectral range, we used the
equivalent width of $\halpha$\ to determine EW([\ion{O}{2}]) = 0.4 EW
($\halpha$) (\markcite{ken92}Kennicutt 1992; \markcite{son94}Songaila et 
al.\ 1994).  EW([\ion{O}{2}]) is shown in Fig.~\ref{fig:10} as a function
of the rest absolute \itk\ magnitude and of redshift $z$; this is perhaps
the most striking single result of the present paper.

At low redshifts (lower left panel) very few galaxies have very strong
[\ion{O}{2}] lines.  Roughly speaking an EW([\ion{O}{2}]) =
$25\ang$\ separates galaxies undergoing rapid formation (mass doubling in
less than $10^{10}~{\rm yr}$) from those which are undergoing low rates of
star formation (\markcite{ken92}Kennicutt 1992), and in this panel only a
small number of very low-mass galaxies fall in the fast-formation category.
As we move to higher redshift, we see progressively more massive galaxies
with rapid formation, until at $z > 1$\ we see $\mkstar \sim -25$\ galaxies
falling in this category (Cowie et al.\markcite{cow95a} 1995a).  Because of
the bias against the spectroscopic identification of red objects the
higher-$z$\ panels may be missing more quiescent objects but the point here
is the presence of many massive rapidly star-forming objects --- a type of
galaxy which is not seen at the present time.  This effect is clearly seen
in other samples:  in particular, \markcite{ell96}Ellis et al.\ (1996) have
shown that the luminosity function of strong [\ion{O}{2}] emitting galaxies
evolves rapidly with redshift whereas that of the weak [\ion{O}{2}] emitters
is nearly invariant.

We can also see the same result independently from the rest-frame
ultraviolet--infrared colors, which provide an independent measure of the star
formation rate in the galaxies that is more direct than the [\ion{O}{2}]
emission (\markcite{ken92}Kennicutt 1992).  We have measured the rest-frame 
$3500\ang$\ $AB$\ magnitude by assigning an approximate type to each galaxy
and using this to interpolate between the observed \itb\ and
\iti\ magnitudes and to extrapolate from \itk\ to obtain the rest-frame
$21000\ang$\ $AB$\ magnitude.

The type was assigned from the galaxy's \imk\ color and redshift by 
interpolating between the predicted colors of Coleman, Wu, \& Weedman 
\markcite{col80}(1980) models at that redshift.  The corresponding spectral
energy distribution was then used to interpolate between the \itb\ and
\iti\ magnitudes.  However, it is important to emphasize that the 
assigned rest-frame \itu\ magnitude is quite insensitive to this procedure
since the interpolation is relatively small and the procedure is, of course,
exact at $z=0.3$ where the observed \itb\ band corresponds to the rest-frame
3500\ang, and at $z=1.4$, where the observed \iti\ magnitude corresponds to 
rest-frame 3500\ang.  To show this insensitivity we have compared the rest-frame
3500\ang\ magnitude computed with the type-interpolation procedure to a
simple logarithmic interpolation where the rest-frame 2500\ang\ magnitude
is obtained as 
\begin{equation}
   (3500)_{AB}\ =\ B_{AB}\ +\ {\displaystyle{\log_{10}\left( 3500\,(1+z)\over 
    4500\right )}\over {\displaystyle{\log_{10}\left( 8400\over 
    4500\right ) }}}\ (I_{AB} - B_{AB})\ .\label{eq:5}
\end{equation}

The deviation between these two methods of assigning the rest-frame \itu\
magnitude reaches a maximum of only 0.3 mags for any galaxy, which provides
a very extreme upper bound to the uncertainty in the 3500\ang\ rest-frame
magnitude introduced by the interpolation.

Fig.~\ref{fig:11} shows the comparison between the [\ion{O}{2}] equivalent
widths and the rest-frame ultraviolet--infrared colors as a
function of redshift and absolute \itk\ magnitude.  At low redshifts almost
all galaxies are quiescent in both diagnostics.  At $z = 0.2- 0.4$, there are
a number of low mass galaxies with large EW([\ion{O}{2}]) and $(3500 - K)_{AB}
< 1.3$\ which are rapid star formers, but more massive galaxies are quiescent.
Finally at higher redshifts massive galaxies are seen in the rapid formation
region.  We can summarize these figures as follows.  It appears that at the
present time there is almost no galaxy formation, but that as recently as $z
\sim 0.2$, low-mass ($\mkstar = -22$) galaxies were forming; progressively
higher galaxy masses are seen in formation at higher redshifts, to the point
where near-$\lstar$\ ($\mkstar = -24$\ to $-25$) galaxies are seen undergoing
rapid star formation at $z > 1$.

\section{The \itb\ Sample}

\subsection{$B - z$\ relation}

The redshift sample was extended to $B = 24.5$\ in both fields by observing
the additional objects not previously included in the $K = 20$\ sample
(Tables \ref{tbl-1a} and \ref{tbl-1b}).  Essentially all the $B =
24.5$\ objects in SSA~22 were observed but in SSA~13 the sample is completely
observed only to $B = 24$\ (Tables \ref{tbl-1a} and \ref{tbl-1b}).  We have
therefore defined the sample as all objects with $B \le 24.5$\ in SSA~22 and
$B < 24$\ in SSA~13.  The total sample comprises 203 objects and is quite
complete ($>$~85\%) even at $B = 24 - 24.5$\ (Fig.~\ref{fig:3}).

The magnitude-redshift relation for the galaxies in the sample is shown in
Fig.~\ref{fig:12}, where we distinguish the SSA~13 members (filled diamonds)
from those in SSA~22 (filled squares).  Unidentified objects are shown as
diagonal crosses at a nominal redshift of 1.75 for SSA~22 and as upright
crosses (pluses) at a nominal redshift of 1.85 for SSA~13.  We also show as
the large pluses the median redshifts for objects in 0.5 mag bins with $\pm
1~\sigma$\ errors computed using the median sign method, again placing
unidentified objects above or below the median to estimate the maximum error.  
For comparison we also show (solid line) the predicted median
redshift-magnitude relation computed for a no-luminosity-evolution model
with the \itb\ luminosity function of Loveday et al.\markcite{lov92}
(1992), the galaxy type mix of King \& Ellis\markcite{kin85} (1985) and the
K corrections of Fig.~1 of Cowie et al.\markcite{cow94} (1994).  This
prediction agrees well with that of Glazebrook et al.\markcite{gla95a}
(1995a), but is significantly higher than that of Colless et
al.\markcite{col93} (1993) apparently because the analytic approximation to
the K corrections used in the latter paper is poor at higher redshifts.
This no-evolution prediction provides a reasonable fit to the data.  The
dashed line shows the Carlberg\markcite{car92} (1992) merger model, which
slightly under-predicts the median redshifts at fainter \itb\ magnitudes.

Previously published $B = 24$\ samples have either been rather small (Cowie
et al.\markcite{cow91} 1991) or relatively incomplete.  The large $B =
24$\ sample of Glazebrook et al.\markcite{gla95a} (1995a) contained 157
objects of which 84 were galaxies with secure redshifts, 8 were stars, and 2
were quasars -- the remaining 63 objects were unidentified, though a smaller
subsample was much more complete.  The present sample contains a very
comparable number of $B < 24$\ objects but is much more complete than
Glazebrook et al., with 149 of the 156 objects identified.  The redshift
distribution of identified objects in the two samples in the $B = 22.5 -
24$\ range is compared in Fig.~\ref{fig:13}, where the present sample is
shown as the solid histogram and the Glazebrook et al.\markcite{gla95a}
(1995a) sample (hereafter GECBAT) is shown as the dashed histogram.  At $z <
0.7$\ the shapes of the two histograms agree well but the present data shows
an extended tail beyond this redshift which is not seen in the GECBAT data.

Most of the objects in the high redshift tail are blue ($B - I$) galaxies
with extremely strong [\ion{O}{2}] emission.  (A number of such objects are
shown in Figs.~\ref{fig:2a} and \ref{fig:2b} and further examples may be
found in Fig.~1 of Cowie et al.\markcite{cow95a} (1995a), in Lilly et
al.\markcite{lil95a} (1995a), and in LeF\`evre et al.\markcite{lef94}
(1994).)  This is illustrated in Fig.~\ref{fig:14}, where we show the the
histogram of the redshift distribution cut by the color criterion $(B - I) =
1.6$.  The blue population divides neatly into a low redshift $z \sim
0.25$\ population and a high redshift $z > 0.8$\ population, while the red
population lies at intermediate redshifts.  This allows us to understand the
difference between GECBAT and the present data.  Most of the unidentified
objects in GECBAT correspond to the blue population (cf.\ GECBAT, Fig.~3)
and, given that $z \sim 0.2$\ blue galaxies have strong emission lines in the
observed optical and are easily identified, the bulk of these missing objects
will be drawn from the high-$z$\ tail.  GECBAT's upper wavelength cutoff was
around $8000\ang$\ so that for objects with $z > 1.1$, the [\ion{O}{2}] line
is outside their spectral window and such objects would not have been
identified.  GECBAT argued that such objects would have been present in the
Colless et al.\markcite{col93} (1993) sample of blue-selected galaxies, where
7 of 13 galaxies with $(B - I) < 1.6$\ were identified and found to be at
modest redshift (all of the $(B - I) < 1$\ objects being so identified).  The
Colless et al.\ galaxies are mostly drawn from brighter magnitudes where the
preponderance of blue objects may be more weighted to low redshifts
(cf.\ Sec.\ \ref{subsec:bcolors-z}), but this sample is also sufficiently
small so that the redshift distribution of its blue objects is not
inconsistent with the result from the present sample, which is that 11 of 36
objects with $(B - I) < 1.6$\ are low-$z$\ objects and 25 of the 36 are at
high $z$.  We therefore conclude that the present data are indeed consistent
with previous results when systematic effects in the incompleteness are
allowed for.

\subsection{Local faint-end blue luminosity function}\label{subsec:localblum}

Shanks\markcite{sha90} (1990) suggested that there might be a steep rise
in the local faint-end blue luminosity function, and Koo and his
collaborators have attempted to provide a self-consistent fit to the number
counts, colors and \itb\ band redshift information by including such a
steeply rising population of blue dwarf galaxies at the faintest
\itb\ magnitudes ($M_{\rm B} \le -16.75$).  Glazebrook et
al.\markcite{gla95a} (1995a) rejected earlier versions of this model based
on their $B = 24$\ sample.  Here we shall consider the most recent version
of this model described in Gronwall \& Koo\markcite{gro95} (1995).
Because of the large number of dwarfs, these models predict that in $B =
24$\ samples we should see a substantial number of galaxies at $z \le
0.1$.  Fig.~\ref{fig:13} shows this for the Gronwall \& Koo\markcite{gro95}
(1995) model, where it is compared with the Glazebrook et al.\ sample and
our own.  We find only one $B \le 24$\ galaxy with $z < 0.1$\ while
Glazebrook et al.\ find 2 --- considerably lower than this model predicts.

To quantitatively test the model, we predicted the expected
number of $B < 24$, $z < 0.1$\ objects in the two areas, using the luminosity
function of objects tabulated in Gronwall \& Koo's\markcite{gro95} (1995)
paper.  We obtained a predicted value of 8 in the current sample and 12 in
the Glazebrook et al.\ sample.  Either sample rejects the Gronwall-Koo
prediction at a high confidence level while in the combined sample the
probability of the Gronwall-Koo luminosity function being acceptable is only
$3 \times 10^{-6}$\ (3 objects observed versus 20 predicted).

We next directly constructed the low-end \itb-band luminosity function
($M_{\rm B} < -20$) using the $1/V$\ method with the combined $z < 0.2$\
sample.  Because almost all of these objects are extremely blue, there is only
a very small K correction over this redshift interval.  The results are shown
in Fig.~\ref{fig:15}, where they are compared with the Loveday et
al.\markcite{lov92} (1992) luminosity function (dashed line) and with the
Gronwall-Koo function (dotted line).  The data are marginally consistent with
a smooth extension of the Loveday et al.\ function but are better fit by a
shallow rise (\markcite{eal93}Eales 1993).  However, they are not consistent
with the Gronwall-Koo function.

\subsection{Colors and redshifts versus \itb\ magnitude}\label{subsec:bcolors-z}

In Fig.~\ref{fig:16}, we show the distribution of redshift versus
\bmi\ color in various \itb\ magnitude slices.  Objects with red
\bmi\ colors ($\gtrsim 1.6$) lie at intermediate redshifts and show a fairly
smooth increase in median redshift with increasing magnitude.  However, the
redshift distribution of $(B - I) < 1.6$\ objects is more complex.  At $B <
23$\ such objects lie primarily at $z \sim 0.25$\ but at $B \sim 24$\ most
lie at $z \gtrsim 0.8$\ with a fraction at $z \sim 0.25$.  This extended
distribution can also be clearly seen in Fig.~\ref{fig:14}.  Most of the
unidentified objects also lie in this color range and may correspond to
objects where the [\ion{O}{2}] 3727 line was moved out of the observed
spectral range ($z \gtrsim 1.6$).  In Fig.~\ref{fig:17} we show the position
of the galaxies as a function of redshift in the $(B - I)$ versus $(I -
K)$\ color-color plane, which also illustrates that most of the unidentified
objects (open diamonds) lie at the same colors as the $z > 1$\ galaxies
(filled diamonds).

\section{The \iti\ Sample}

\subsection{$I - z$}

The $I - z$\ relation is given in Fig.~\ref{fig:18}, with filled squares
showing the SSA~22 field to $I = 23$\ and diamonds SSA~13 to $I = 22.5$.
The small number of objects which were not observed or not identified are
shown as plus signs for SSA~22 and crosses for SSA~13.  The large upright
crosses show the median redshifts with 1 $\sigma$ error bars computed using
the median sign method and allowing unidentified or unobserved objects to be
at either high or low redshift.

The CFRS survey has produced a very large, spatially sparse sample of
galaxies to $I = 22$\ whereas the present sample contains all galaxies in
our areas to this magnitude.  The areal densities of objects versus
redshift in the two samples for $I = 17 - 22$\ are compared in the upper
panel of Fig.~\ref{fig:19} and agree strikingly well in shape, and more
approximately in normalization, with the present sample -- being about 20\%
below the CFRS value.  Lilly et al.\markcite{lil95b} (1995b) have used the
CFRS sample to predict the redshift distribution at fainter magnitudes and
we compare this (lower panel: dotted line) with the observed distribution
at $I = 22-23$.  This CFRS extrapolation has a similar shape to the
observed $I = 22-23$\ sample, but is slightly overweighted to high redshift
objects. 

The CFRS includes 28 objects with $I < 22$\ that lie in the present SSA~22
field region.  All of these objects are contained in the present catalog and
the \iti\ magnitudes are in good agreement, with $\langle I - I^{\rm
CFRS}_{AB} \rangle = 0.5 \pm 0.25$, as expected.  The redshifts given in the
two surveys agree in all cases where a CFRS redshift was obtained.  Even in
the one case where an object was excluded from the CFRS analysis with a low
confidence (22.1210 with a value of 1) the CFRS redshift agrees with the
present value.

\subsection{The high-$z$\ rest-frame \itb-band luminosity function}

Following Lilly et al.\markcite{lil95b} (1995b), we have used the \iti-band
sample to create the rest-frame \itb-band luminosity function directly at $z
\sim 0.8$, where the observed \iti\ corresponds to rest-frame \itb.  At this
redshift the rest-frame absolute \itb\ magnitude $M_{\rm B}$\ is given by
\begin{equation}
   M_{\rm B} = I -dm +  1.4 \label{eq:6}
\end{equation}
where $dm$ is the distance modulus and the final term is the K correction
allowing for the relative wavelengths and zero points.  For neighboring
redshifts, there is a small type-dependent K correction to equation (\ref{eq:6})
which we have assigned using an approximate type derived from the color and
redshift, following the procedures outlined in Sec.\ \ref{subsec:col-line-evol}.
However, the results are essentially unaffected by this small correction.

The rest-frame \itb-band Schechter function, computed from the \iti\ sample
using the $1/V$\ method in the redshift interval $z = 0.6-1.0$, is shown in
the upper panel of Fig.~\ref{fig:20} as the large open boxes.  Downward
pointing arrows denote a $1~\sigma$\ upper limit while $\pm 1~\sigma$\ error
bars are superimposed on the symbols, assuming Poisson errors based on the
number of objects in each bin.  The solid line shows a fit to the points with
the assumption that the luminosity function has the same shape as the local
value ($\mbstar = -21.0 + 5 \log_{10}\,h_{50}$, $\alpha = -0.97$) derived by
Loveday et al.\markcite{lov92} (1992).  This provides an acceptable fit to
the data with $\phistar = 4.1 \times 10^{-3}\,h_{50}^3~{\rm
Mpc}^{-3}$\ compared to the local value of $1.8\times 10^{-3}\,h_{50}^3~{\rm
Mpc}^{-3}$ of Loveday et al.\ which is shown as the dashed line.  The rise in
normalization is similar to that derived by Eales\markcite{eal93} (1993).

For comparison we have also shown the corresponding luminosity function
derived by Lilly et al.\markcite{lil95b} (1995b) for $z = 0.5-0.7$\ (open
triangles) and $z = 0.7-1.0$\ (open diamonds).  These fall somewhat above the
present data, reflecting the slightly higher normalization in the histogram
of Fig.~\ref{fig:19}.  We have also shown (filled squares) the local
faint-end \itb-band luminosity function derived in Sec.\ \ref{subsec:localblum}.

The lower panel also shows the relative rest-frame \itk\ band luminosity
functions both locally and at $z = 0.6 - 1$.  The infrared normalization
rises by a considerably smaller amount (1.25).  In combination the two
results show that the rest-frame light is considerably bluer at $z = 1$\ (by
about 0.6 mag in the rest frame $(B - K)$), which is consistent with the
discussion of the colors and line strengths of the \itk\ sample in
Sec.\ \ref{subsec:col-line-evol}.  However, Fig.~\ref{fig:20} makes the
important point that at these redshifts the evolution appears to be primarily
in the $L < \lstar$\ galaxies and not to be affecting the colors or
luminosities of the near-$\lstar$\ galaxies; this reflects the differential
luminosity evolution we shall discuss in the next section.

\section{Interpretation}\label{sec:interpret}

\subsection{$\mk - t_d - z$}

Each galaxy in the sample can be roughly characterized by three quantities,
$z$\ the redshift, $\mk$, which for older galaxies is a rough measure of the
total mass in stars, and either an emission-line equivalent width 
[EW([\ion{O}{2}]) or EW($\halpha$)] or a rest-frame ultraviolet color [$\fnu
(3500)/\fnu(21000)\ \equiv\ UK(AB)$], which roughly measures the
rate of star formation relative to the existing stellar mass.  The rest-frame
ultraviolet--infrared color is a more direct measure of the stellar
properties than the emission line strengths (\markcite{ken92}Kennicutt 1992),
and here we use $UK(AB)$\ rather than equivalent widths.  For galaxies with
little internal opacity $UK(AB) \equiv (3500-21000)_{AB} \equiv -2.5
\log_{10} {\fnu (3500)/\fnu(21000)}$\ is a monotonic function of the stellar
mass doubling time, as is illustrated in Fig.~\ref{fig:21}, which is computed
from the models of Bruzual \& Charlot\markcite{bru93} (1993) for a Salpeter
IMF.  Galaxies with $UK(AB) \le 1.3$\ have doubling times less than
$10^{10}$\ years and may be thought of as ``in formation,'' where formation
here means only that the bulk of stars would be made if the process persisted
for a fraction of the local Hubble time, and does not necessarily mean that
there has not been earlier star formation.  Uncertainties in the $UK(AB)$ 
color of 0.3 mags, which represents the maximum deviation between the
two interpolation methods discussed in Sec.\ \ref{subsec:col-line-evol}, would 
introduce uncertainties of about a factor of 2 in
$t_d$ for $t_d \sim \ten{10}$ yrs.  However, a more important source of
uncertainty is internal extinction.  Internal extinction implies that
the doubling time is overestimated using the observed $UK(AB)$.  For a
typical internal extinction of $E(B - V) = 0.2$\ to HII regions in blue
irregulars (\markcite{gal89}Gallagher et al.\ 1989), the intrinsic
$UK(AB)$\ would be roughly a magnitude bluer than the observed value, with a
corresponding reduction in $t_d$.

The position of each galaxy in the three-dimensional $\mk - t_d -z$\ space is
shown in Fig.~\ref{fig:22}, which synthesizes the first main conclusion of
the present paper (see Sec.\ \ref{subsec:col-line-evol} for a more
phenomenological version).  We can summarize the figure as follows: At $z <
0.2$\ there are no forming galaxies above $\mk = -20 + 5
\log_{10}\,h_{50}$\ ($0.01\lstar$).  At $z = 0.2 - 0.4$, we see forming
galaxies at $\mk = -20$\ to $-22.5 + 5\log_{10}\,h_{50}$\ ($0.01 -
0.1\lstar$).  At $z = 0.4 - 0.8$\ forming galaxies are seen which are as
bright as $\mk = -23.5 + 5\log_{10}\,h_{50}$, and at $z = 0.8$\ to $z =
1.6$\ as bright as $\mk \sim -24 + 5\log_{10}\,h_{50}$.

The time interval between the redshift slices is roughly 10\% to 15\% of the
current age of the universe, depending on the assumed geometry, and it is
clear that there is substantial migration in the $t_d - \mk$\ space between
each redshift interval.  As is illustrated in Fig.~\ref{fig:22} for the
Bruzual--Charlot models for various exponential star-formation models ($\tau
= 3~{\rm Gyr}$ -- dotted, $\tau = 7\ {\rm Gyr}$ -- solid, and constant
SFR -- dashed), it is clear that this migration requires that the SFR be
declining faster than a 3~Gyr exponential to move the fast-forming galaxies
away from their initial locus.  This result may be weakened somewhat if the
evolution is accompanied by an increase in internal extinction.

Fig.~\ref{fig:22} leaves open the question of the final location of the
galaxies which could lie at magnitudes only slightly fainter than their
initial $\mk$\ ($\tau = 3\ {\rm Gyr}$\ exponential SFR) in which case we
are seeing the formation of the massive end of the Schechter function, or
at several magnitudes fainter ($10^8$\ yr burst) in galaxies that were only
transiently luminous. This question is best addressed in terms of the
distribution and evolution of the \itk\ light density and the UV light
density production, to which we now turn.

\subsection{\itk-band light density evolution}\label{subsec:klightevol}

The rest-frame \itk-band light density provides our best available tracer of
the baryonic mass which has formed into stars.  The present day \itk-band
light density of $2.5 \times 10^8\lsun$\ to $\mk = -20 + 5
\log_{10}\,h_{50}$\ ($L = 0.01\lstar$), as measured in Fig.~\ref{fig:6a},
translates to a stellar baryonic mass density of $2 \times
10^8\,h_{50}\msun\ {\rm Mpc}^{-3}$\ ($\Omega = 3\times 10^{-3}\,h_{50}^{-1}$)
for a mass-to-\itk-light ratio of 0.8 in solar units.  
As is obvious from the shape of the \itk-band Schechter function, more
than 85\% of the light lies in luminous galaxies ($\mk = (-22 \rightarrow
-26) + 5\log_{10}\,h_{50}$).

In Fig.~\ref{fig:23} we have constructed the \itk-band cumulative luminosity
density for the $K = 20$\ selected sample for $z = 0.0 - 0.3$\ and $z = 0.3 -
1$, as was done in Fig.~\ref{fig:6a}, but have now split it into $t_d <
10^{10}~{\rm yr}$\ and $t_d > 10^{10}~{\rm yr}$\ samples ($UK(AB) < 1.3$\ and
$UK (AB) > 1.3$).  In each case the old (red) sample is shown by crosses and
the young (blue) sample by boxes.  In the lower redshift bin 5\% of the light
to $\mk = -20 + 5\log_{10}\,h_{50}$\ comes from the young objects and these
are all faint ($\mk$ fainter than $-22 + 5\log_{10}\,h_{50}$).  However, in
the higher redshift bin the fraction of light in forming objects is
considerably higher (10\% to $\mk = -21$) and also extends to considerably
brighter magnitudes.  For comparison we have shown as a solid line the shape
of the cumulative luminosity density in the lower redshift interval
renormalized to the higher redshift curve, which shows clearly that the
extension to brighter magnitudes is not simply an effect of the higher
normalization but represents a brightening of the luminosity of forming
galaxies at higher redshift.  

This relatively rapid evolution in blue-selected versus red-selected samples 
has also been emphasized by Lilly et al.\markcite{lil95b} (1995b) based on 
the CFRS \iti-band sample and the same effect has been found by Ellis et al.\ 
(1996)\markcite{ell96} using line diagnostics.  Ellis et al.\ find that the 
luminosity function of strong [\ion{O}{2}] line emitters evolves rapidly but 
that of weak [\ion{O}{2}] emitters does not.  It is probable that the rapid 
increase in the fraction of galaxies with unusual morphologies at increasing
$z$ (Glazebrook et al.\ 1995b\markcite{gla95b}, Cowie et al.\ 1995b%
\markcite{cow95b}, Driver et al.\ 1995\markcite{dri95}) also reflects
this underlying evolution.

The luminosity function of the old (red) galaxies shows very little 
evolution of its shape over this redshift interval, and with the exclusion 
of the young objects which are much more rapidly evolving, it is found that 
the red-selected sample evolves less in light density than the total sample.  
From Fig.~\ref{fig:23}, $\lambda_{\rm
K}$\ to $\mk = -21$\ has changed from $3.3 \times 10^8\lsun\ {\rm
Mpc}^{-3}$\ for $z = 0.3$\ to 1, to $2.5 \times 10^8\lsun\ {\rm
Mpc}^{-3}$\ for $z = 0 - 0.3$, for $q_0 = 0.5$, while for $q_0 = 0.02$\ this
becomes $2.6 \times 10^8\lsun\ {\rm Mpc}^{-3}$\ at the higher redshift
interval and $2.4 \times 10^8\lsun\ {\rm Mpc}^{-3}$\ at the lower.

Fig.~\ref{fig:24} shows the cumulative \itk-band light density for the blue
objects over a wider range of redshifts where we now use a $K = 20.5$\ sample,
since to this magnitude all objects with $UK(AB) < 1.3$\ are contained in the
combined \itk, \iti, and \itb\ sample.  Fig.~\ref{fig:24} shows the smooth
increase in luminosity of the most luminous forming galaxies with increasing
redshift.  It is this smooth change of the maximum ``forming'' luminosity
with redshift which, when combined with the distance modulus and the
rest-frame blue colors of the ``forming'' galaxies, results in a near caustic
where the apparent magnitude at which the forming galaxies enter the samples
is almost the same at all redshifts --- roughly $B \geq 23$ or $K \geq 20$.
This is illustrated at a more detailed level in Fig.~\ref{fig:25} where we
show the ``forming'' galaxies selected either by UV color or [\ion{O}{2}]
equivalent width (filled squares) in $z-B$ space superimposed over all $B <
24.5$ galaxies (pluses).  The first ``forming'' galaxies to enter the sample
lie at around $B\sim22$ and $z\sim0.3$, and it is this minimum which resulted
in the `small blue galaxies' being the first detected population of forming
galaxies (Cowie et al.\markcite{cow91} 1991, Glazebrook et
al.\markcite{gla95a} 1995a), but beyond $B=23$ forming galaxies enter at
essentially all redshifts to at least $z=1.7$ (Fig.~\ref{fig:25}).  It is the
rather remarkable cancellation of the ``forming'' $\mk$\ and the distance
modulus which results in the excess in the \itb-band counts at $B \gtrsim 23$
and the rapid blueing which sets in at the faint \itb\ and \itk\ magnitudes.

As was shown in Sec.\ \ref{subsec:col-line-evol} the \itk\ luminosity
density rises with increasing redshift at intermediate redshifts but then
begins to drop at higher redshift ($z \gtrsim 0.6$).  
This effect can be easily understood if the $\mk - t_d - z$\ diagrams
of the previous subsection are interpreted as showing the formation of
present-day luminous galaxies (exponential star-forming galaxies) since in
this case much of the present-day \itk\ luminosity density, which, as we
have shown above, arises primarily at $\mk < -23 + 5\log_{10}\,h_{50}$, was
being assembled at the earlier redshifts.  The \itk\ luminosity density
peaks when the more massive galaxies cease to form, and only then begins to
passively decline.  By contrast, if the rapid formers were bursts in smaller
objects, the bulk of the \itk\ light would already be present in invariant
objects at these redshifts and we would expect to see a smooth increase in
$\lambda_{\rm K}$\ with increasing $z$\ from the passive evolution.

\subsection{The universal rate of star formation.}

The \itk\ light density discussed in the previous subsection is an integral
measure of the accumulated light density as a function of redshift and so its
evolution provides a history of universal star formation.  An alternative
approach is to look at the rest-frame ultraviolet light density, which
measures the rate of massive star formation and so, for an invariant stellar
mass function, the rate of star formation.  The ultraviolet light density
should therefore roughly measure the rate of accumulation of the light
density which, when convolved through the transfer function from stellar
evolution, would predict the \itk\ light density evolution.  It therefore
provides an independent check of the results of the previous subsection.

Cowie\markcite{cow88} (1988) first suggested that a very model-invariant way
to describe this was by comparing the extragalactic background surface
brightness at these frequencies with the production of metals, since both
relate to the same massive stars and so avoid the uncertainties in the
stellar mass function.  Songaila et al.\markcite{son90} (1990) give this
relation as
\begin{equation}
   S_{\nu} = 3.6 \times 10^{-25} \left ( {{\rho Z} \over {10^{-34}~{\rm g\ 
   cm^{-3}}}}\right )\quad {\rm ergs\ cm^{-2}\ s^{-1}\ Hz^{-1}\ deg^{-2}}
   \label{eq:7}
\end{equation}
where $S_{\nu}$\ is the present sky surface brightness at the rest-frame
ultraviolet
frequency $\nu$\ and $\rho Z$\ is the present mass density of metals in the
universe.  This equation is independent of the cosmological parameters.  For
comparison purposes we shall assume a local $\rho Z = (2\rightarrow 6) \times
10^{-34}\,h_{50}^2~{\rm gm\ cm^{-3}}$\ following the various estimates given
in Cowie\markcite{cow88} (1988) and Songaila et al.\markcite{son90} (1990).

Fig.~\ref{fig:26} shows $S_{\nu}$\ for objects satisfying $UK(AB) <
1.3$\ divided by redshift bin compared with the expected sky surface brightness
which should be produced by all galaxy formation, the range of which is shown 
by the dotted lines.  The upper right panel shows the rest-frame ultraviolet
surface brightness of all galaxies in the samples irrespective of their
colors, while the lower right panel shows that of only those objects
whose rest-frame ultraviolet--infrared colors imply rapid formation.
Finally, in the two left-hand panels (solid lines) we show the extragalactic 
background light (EBL) for the rapidly forming galaxies of the lower right 
panel, cut at an absolute $3500\ang$\ magnitude of $-20.5 +
5\log_{10}\,h_{50}$.  From equation~(4) of Cowie\markcite{cow88} (1988),
this roughly corresponds to an $\dot M$\ of $30\msun~{\rm yr}^{-1}$\ for a
Salpeter IMF, and so is approximately the luminosity which will form a
$10^{11}\msun$\ galaxy in $3\times10^9~{\rm yr}$.  The two left-hand panels 
again illustrate the point that larger galaxies form earlier than smaller
ones, but also illustrate the second point, that we are seeing a large
fraction (10--20\%) of the light needed to make all observed galaxies in
the rapid formers.  In fact, this conclusion is substantially understated
by the solid lines because of internal extinction.  If we allow for one
magnitude of internal extinction in the galaxies, the solid line of massive
forming galaxies in the upper left panel moves to the dashed line and
essentially all of galaxy formation can be accounted for.

The right-hand lower panel shows the total observed light in all forming
galaxies, again assuming no extinction, and in the upper right we compare this
with the light of all observed galaxies.  Again we see that, with even a small
amount of extinction and any level of incompleteness at the high-$z$\ end, we
are seeing the bulk of galaxy formation.  The upper panel also makes the
point that roughly half of the star formation occurs in the fast-forming
galaxies, while the other half occurs as ongoing star formation in older
galaxies, as would be expected in exponential star-forming models, but not in
burst models.

\section{Conclusions}

In this paper we have presented evidence that some substantial fraction of all
galaxy formation occurred between $z = 0.8$\ and $z = 1.6$, and also that
galaxy formation took place in ``downsizing'', with more massive
galaxies forming at higher redshift.  The late galaxy formation accounts for
the rapid evolution in galaxy colors, while the differential evolution ---
with the most massive galaxies forming earliest --- results in relatively
little evolution in $\kstar$\ and $\bstar$\ between $z = 0$\ and $z = 1$\
and so results in the galaxy redshift distribution being well described by
no-evolution models at $z < 1$.

The present data fit well with recent studies of the evolution of the
intergalactic gas.  In particular, it now appears that the $\Omega$\ in damped
$\lal$\ absorption systems seen in quasar spectra was relatively constant from
$z = 2$\ to 5, with a value of around $2.5\times 10^{-3}\,h_{50}$, very
similar to that in present-day galaxies (Sec.\ \ref{subsec:klightevol}), but
drops rapidly at $z < 2$\ (\markcite{sto95}Storrie-Lombardi et al.\ 1995)
presumably as a consequence of the galaxy formation discussed here.  The onset
of rapid galaxy formation at these redshifts also allows us to understand the
low metallicity in most $z \gtrsim 2$\ damped $\lal$\ systems
(\markcite{pet94}Pettini et al.\ 1994) and the correspondingly required cosmic
chemical evolution between $z = 2$\ and $z = 0$\ (\markcite{pei95}Pei \& Fall
1995).

The ``downsizing'' is a more remarkable result.  It allows a very natural
interpretation of the data at the expense of a remarkably regular fall of
the maximum luminosity of ``forming'' galaxies with redshift.  It is
possible to envisage theoretical models which utilize feedback through the
intergalactic gas to achieve this type of result as, for example, if the
ongoing galaxy formation increased the IGM pressure and so reduced the
characteristic ``forming'' mass, but it remains to be seen whether a fully
self-consistent description can be constructed which also fits the quasar
absorption line data.

\smallskip
\acknowledgments
We would like to thank Karl Glazebrook, St\'ephane Charlot, Tom Shanks, Caryl
Gronwall, Tom Broadhurst, and most particularly the referee, Richard Ellis,
for their many helpful comments on an earlier version of this paper. We are
grateful to Tom Bida, Peter Gillingham, Joel Aycock, Teresa Chelminiak,
Barabara Schaefer, and Wayne Wack for their extensive help in obtaining the
observations with the Keck LRIS spectrograph.  This work was partly based on
observations obtained with the NASA/ESA {\it Hubble Space Telescope} and was
supported by the State of Hawaii and by NASA through grants number
GO-5399.01-93A, GO-5922.01-94A, and GO-6626.01-95A from the Space Telescope
Science Institute, which is operated by AURA, Inc., under NASA contract
NAS5-26555.

\smallskip
\acknowledgments

\newpage

\figurenum{1a}
\figcaption[cowie.fig1a.ps]
{\label{fig:1a}$I$-band finding chart for the objects in Table
\protect{\ref{tbl-1a}}. Each object is circled and labelled with the
running object number given in column~1 of the table.}

\figurenum{1b}
\figcaption[cowie.fig1b.ps]
{\label{fig:1b}The corresponding $I$-band finding chart for Table
\protect{\ref{tbl-1b}}.}

\figurenum{2a}
\figcaption[cowie.fig2a.ps]
{\label{fig:2a}Spectra of every 10th object in Table \protect{\ref{tbl-1a}},
excluding stars and unidentified objects.  The spectra are shown in the rest
frame corresponding to the redshift given at the top of each panel, and are
approximately $f_{\nu}$\ in arbitrary  units.  The shaded regions mark the
positions of the stronger night sky lines where residuals may be present in
the spectra, and the positions of the atmospheric bands.  The $K$\ and
$B$\ magnitudes are shown at the lower right, and the identifying number from
Table \protect{\ref{tbl-1a}} (column 1) at the lower right.}

\figurenum{2b}
\figcaption[cowie.fig2b.ps]
{\label{fig:2b}Spectra of every 10th object in Table \protect{\ref{tbl-1b}},
excluding stars and unidentified objects.  The spectra are shown in the rest
frame corresponding to the redshift given at the top of each panel, and are
approximately $f_{\nu}$\ in arbitrary  units.  The shaded regions mark the
positions of the stronger night sky lines where residuals may be present in
the spectra, and the positions of the atmospheric bands.  The $K$\ and
$B$\ magnitudes are shown at the lower right, and the identifying number from
Table \protect{\ref{tbl-1b}} (column 1) at the lower right.}

\figurenum{3}
\figcaption[cowie.fig3.ps]
{\label{fig:3}Completeness of the sample in each color as a function of
magnitude.  The completeness is defined here as the ratio of the total number
of identified objects (stars and galaxies) to the total number of objects in
the catalog (whether observed or not).  Data from Table
\protect{\ref{tbl-1a}} (SSA~13 field) are only included for $I < 22.5$\ and
$B< 24$.  The total number of objects in each color sample is shown at the
lower left.}

\figurenum{4}
\figcaption[cowie.fig4.ps]
{\label{fig:4}Redshift versus $K$\ magnitude for the present sample at $K >
18$\ (filled squares).  At $K \le 18$\ we show the much larger Songaila et
al.\protect{\markcite{son94}} (1994) sample (filled triangles) which contains 
the present areas as a subsample.
Open triangles show the data of Mobasher et
al.\protect{\markcite{mob86}} (1986) and open diamonds the
$40\,h_{50}^{-1}~{\rm kpc}$\ metric magnitudes of Glazebrook et
al.\protect{\markcite{gla95b}} (1995b).  We have also shown the
unidentified or unobserved objects in Tables \protect{\ref{tbl-1a}} and
\protect{\ref{tbl-1b}} as crosses at an arbitrary redshift for each field.
The solid lines show the apparent $K$\ magnitude for an Sb galaxy with $L =
3L_{\ast}$\ (upper) and $L = 0.1L_{\ast}$\ (lower), where we take $M_{\rm
K\ast} = -25.0 +5\log_{10}\,h_{50}$.}

\figurenum{5}
\figcaption[cowie.fig5.ps]
{\label{fig:5}Median redshift as a function of $K$\ magnitude.  The open
squares are from Mobasher et al.\protect{\markcite{mob86}} (1986), the filled
squares from Songaila et al.\protect{\markcite{son94}} (1994), the open
diamonds from Glazebrook et al.\protect{\markcite{gla95b}} (1995b), and the
pluses from the present work.  In each bin the error bars are $\pm
1\sigma$\ based on the median sign method, allowing unidentified or
unobserved objects to be at either high or low redshift.  The solid ($q_0 =
0.5$) and dashed ($q_0 = 0.02$) lines show the predictions for a non-evolving
luminosity function with $M_{{\rm K}{\ast}} = -25.0 +5\log_{10}\,h_{50}$\ and
$\alpha = -1.25$.  The dash-dot line shows a model with mild luminosity
evolution while the dotted line shows the prediction of a merger model
\protect{\markcite{car92}}Carlberg 1992).}

\figurenum{6a}
\figcaption[cowie.fig6a.ps]
{\label{fig:6a}The cumulative comoving rest-frame $K$-band luminosity density
from galaxies more luminous than a given rest-frame absolute $K$\ magnitude.
The pluses correspond to the redshift ranges $z = [0,0.3]$, filled squares to
$z = [0.3,1]$, and filled diamonds to $z = [1,1.6]$.  The left panel shows
$q_0 = 0.5$, the right, $q_0 = 0.02$.  The solid line shows the $z =
[0.3,1]$\ curve renormalized to match the local luminosity density.  Note
that the asymptotic luminosity density has risen by roughly a factor of 1.4
to 1.6 (depending on $q_0$) between the first two redshift intervals (pluses
and filled squares).  The $K$\ magnitude cutoff for the sample does not allow
similar extrapolation for the $z = [1,1.6]$ interval, but the marked
divergence of its curve at luminosities brighter than $M_{\rm K} \approx -25$
and the change in its shape with respect to the cumulative luminosity density
functions of the lower two redshift intervals may be noted.}

\figurenum{6b}
\figcaption[cowie.fig6b.ps]
{\label{fig:6b}The corresponding histogram of the number of objects in each
absolute magnitude bin.  The solid line is for the $z = [0,0.3]$ redshift
interval and the dashed is for $z = [0.3,1]$.  The left panel is for 
$q_0 = 0.5$, the right is for $q_0 = 0.02$.}

\figurenum{7}
\figcaption[cowie.fig7.ps]
{\label{fig:7}$1~\sigma$\ error contours for $K_{\ast}$\ and $\alpha$\ in a
Schechter function fit to the cumulative number distribution of the data at
$z = [0,0.3]$ (solid) and $z = [0.3,1]$ (dashed).  The filled square shows
the best fit at low redshift and the open triangle the best fit at high
redshift, and within the statistical and photometric uncertainties these
describe an invariant, or possibly even declining (depending on $q_0$)
$K$-band luminosity across the spanned redshift interval.}

\figurenum{8}
\figcaption[cowie.fig8.ps]
{\label{fig:8}$K$-band luminosity function as a function of redshift
slice.  The filled squares show only the Hawaii data (with downward
pointing arrowheads indicating 1 $\sigma$ upper limits in bins where no
objects were detected), while the open squares also include the Mobasher
et al.\protect{\markcite{mob93}} (1993) and Glazebrook et
al.\protect{\markcite{gla95b}} (1995b) data. The solid line shows the best
fit Schechter function with $M_{\rm K\ast} = -25.0$, $\alpha =
-1.25$\ (Hawaii data: solid; all data:  dashed) while the dotted line
shows the function for $\phi_{\ast} = 10^{-3}~{\rm Mpc}^{-3}$.}

\figurenum{9a}
\figcaption[cowie.fig9a.ps]
{\label{fig:9a}$(I - K)$ versus $K$\ for the present sample (panel~$a$).
All objects are shown with $\pm 1~\sigma$\ error bars in $I$.  The dotted
lines show the magnitude limits of the current spectroscopic samples while
the solid line shows the expected $(I - K)$ color of a $z = 1.5$ Sb galaxy,
and the dashed line indicates flat $f_{\nu}$.  Identified objects are shown
as asterisks (stars), filled squares (galaxies with $z < 1$), and boxed
filled squares (galaxies with $z > 1$), with unidentified objects designated
with pluses.}

\figurenum{9b}
\figcaption[cowie.fig9b.ps]
{\label{fig:9b}The corresponding $(I - K)$ versus $K$\ plot for
a very deep sub-sample (panel~$b$).  Although the incompleteness in
the $K$-selected sample is primarily in objects with redder $(I - K)$ colors
[panel a], blue objects dominate the \itk\ counts, and the great bulk of
$K$-selected objects are extremely blue in $(I - K)$ [panel b].}

\figurenum{10}
\figcaption[cowie.fig10.ps]
{\label{fig:10}Rest-frame [O II] equivalent width versus absolute rest $K$
magnitude and redshift for the $K < 20$\ sample.  In the lowest redshift
interval (lower left panel) very few galaxies have strong [O II] lines
or are undergoing rapid star formation (EW([O II]) $\gtrsim25$\ang).
At higher redshifts, progressively more massive galaxies are undergoing
rapid formation, until at $z>1$ the locus of rapidly forming galaxies
reaches $\mkstar\sim -25$.}

\figurenum{11}
\figcaption[cowie.fig11.ps]
{\label{fig:11}Comparison of the rest-frame equivalent width of [O II] 3727
in angstroms with the rest-frame $(3500 - K)$\ color in $AB$\ magnitudes for
various redshift slices.  Pluses are galaxies with $M_{\rm K} < -22$, filled
squares superposed on pluses mark more luminous galaxies.  Galaxies with
strong star formation should have EW([O II]) $\gtrsim 25~{\rm\AA}$\ and
$(3500 - K)_{AB} \lesssim 1.3$\ (dashed lines).}

\figurenum{12}
\figcaption[cowie.fig12.ps]
{\label{fig:12}Redshift versus apparent $B$\ magnitude.  Diamonds = SSA~13
field to $B = 24$, squares = SSA~22 field to $B = 24.5$.  The crosses and
pluses show unobserved or unidentified objects at a nominal redshift.  The
large crosses show the median redshifts with $\pm 1~\sigma$\ error bars in
half-magnitude intervals compared to a no-luminosity evolution model
described in the text (solid line) and a merger model
(\protect{\markcite{car92}}Carlberg 1992) shown as a dashed line.}

\figurenum{13}
\figcaption[cowie.fig13.ps]
{\label{fig:13}Histogram of the redshift distribution of $B = 22.5$\ to $B
= 24$\ galaxies in the present sample (solid histogram) compared with that
of Glazebrook et al.\protect{\markcite{gla95a}} (1995a) (dashed
histogram).  Also shown are the predictions of a luminosity evolution model
(solid line), a merger model (Broadhurst et al.\protect{\markcite{bro92}}
1992; dashed line) and a steep dwarf luminosity function model (Gronwall \&
Koo\protect{\markcite{gro95}} 1995; dotted line).  The latter model
predicts a substantial number of faint blue objects at $z < 0.1$,
corresponding to a large population of dwarf galaxies, which are not seen
in the present sample.}

\figurenum{14}
\figcaption[cowie.fig14.ps]
{\label{fig:14}$B = 22.5$\ to $B = 24$\ redshift histogram divided by $(B -
I)$\ color.  (Solid line: $(B - I) < 1.6$, dashed line: $(B - I) > 1.6$.)
Most of the objects in the high-redshift tail $(z \gtrsim 0.8)$ are blue [$(B
- I) < 1.6$] and show strong [O II] emission.  The $(B - I)$ color criterion
is shown to divide the blue population into low-redshift ($z\sim 0.25$) and
high-redshift $(z \gtrsim 0.8)$ components, with the red population lying at
intermediate redshifts.}

\figurenum{15}
\figcaption[cowie.fig15.ps]
{\label{fig:15}Local faint-end $B$\ luminosity function (filled squares
with $1~\sigma$\ error bars) computed from the present data and Glazebrook
et al.\protect{\markcite{gla95a}} (1995a).  The dashed line shows the
luminosity function determined by Loveday et al.\protect{\markcite{lov92}}
(1992) and the dotted line the function assumed by Gronwall \&
Koo\protect{\markcite{gro95}} (1995).  The data are consistent with a
smooth extension of the Loveday et al.\ function to brighter magnitudes or
with a shallow rise (\protect{\markcite{eal93}}Eales 1993), but are not
consistent with either the Gronwall-Koo function or the steep rise given by
Shanks (\protect{\markcite{sha90}}1990) at the bright end.}

\figurenum{16}
\figcaption[cowie.fig16.ps]
{\label{fig:16}Redshift versus $(B - I)$\ color and $B$\ magnitude.  Squares
are from the SSA~22 field and diamonds from the SSA~13 field.  Pluses show
unidentified objects at a nominal redshift.  Objects with red $(B - I)$\
colors ($ \gtrsim 1.6$) lie at intermediate redshifts.  The bimodal
distribution of blue [$(B - I) < 1.6$] objects into low- ($z \sim 0.25$) and
high-redshift ($z \gtrsim 0.8$) populations shown in
Fig.~\protect{\ref{fig:14}} may be seen in more detail here, and is seen to
onset at $B \sim 23$.  Most of the unidentified objects (pluses) also lie
within this range of blue $(B - I)$\ colors, and many of these may correspond
to objects where the [O II] 3727 line moves out of the observed spectral
range ($z \gtrsim 1.6$).}

\figurenum{17}
\figcaption[cowie.fig17.ps]
{\label{fig:17}Color-color diagram of the $B$\ selected sample showing the
positions of stars (asterisks), $z < 0.3$\ galaxies (filled squares), $0.3 <
z < 1$\ galaxies (pluses), $z > 1$\ galaxies (filled diamonds) and
unidentified objects (open diamonds).  Most of the unidentified objects have
blue $(B - I)$\ colors and redder $(I - K)$\ colors and lie in the portion of
the color-color plane (dashed lines) where most of the $z > 1$\ galaxies
(filled diamonds) also lie.}

\figurenum{18}
\figcaption[cowie.fig18.ps]
{\label{fig:18}Redshift versus $I$\ magnitude with filled squares showing
data for the SSA~22 field ($I < 23$) and filled diamonds the SSA~13 field ($I
< 22.5$).  Pluses and crosses are unidentified or unobserved objects placed
at a nominal redshift.  The large upright crosses show median redshifts with
1 $\sigma$ errors computed using the median sign method.}

\figurenum{19}
\figcaption[cowie.fig19.ps]
{\label{fig:19}(Upper panel)\quad Comparison of the surface density of objects
versus redshift (histogram) in the present sample with $I = 17 - 22$\ with
that in the CFRS (solid line). \quad (Lower panel) \quad Comparison of the
present data with $I = 22 - 23$\ (histogram) versus the no-evolution
prediction from CFRS (Lilly et al.\protect{\markcite{lil95b}} 1995b).}

\figurenum{20}
\figcaption[cowie.fig20.ps]
{\label{fig:20}Comparison of the rest $K$- (lower) and rest $B$-band
luminosity functions at $z = 0 - 0.2$\ (filled squares) and $z = 0.6 -
1$\ (open squares).  In the upper panel we show similar results for the
CFRS at $z = 0.5 - 0.7$\ (open triangles) and $z = 0.7 - 1$\ (open
diamonds).  The dashed line in the upper panel shows the Loveday et
al.\protect{\markcite{lov92}} (1992) Schechter function and the solid line
a Schechter function with the same shape but normalized to the high
redshift data.  The lower panel has similar fits for $\mkstar = -25.0$,
$\alpha = -1.25$, and shows a considerably smaller rise in the infrared
normalization (factor of 1.25) between the low- and high-redshift data
(dashed and solid curves) than is seen for the \itb\ data in the upper
panel.  The data in combination show that the rest-frame light is
considerably bluer at $z\sim1$ (see, e.g., Figs.~\protect{\ref{fig:14}}
and \protect{\ref{fig:16}}), and that at these redshifts the evolution is
primarily in sub-\lstar\ galaxies, while the colors and luminosities of
near-\lstar\ galaxies are not substantially affected.}

\figurenum{21}
\figcaption[cowie.fig21.ps]
{\label{fig:21}Relation between rest-frame $(3500 - K)_{AB}$\ and the 
stellar mass doubling time, computed from the models of Bruzual \&
Charlot\protect{\markcite{bru93}} (1993).  Galaxies with $(3500 - K)_{AB}
\lesssim 1.3$ have doubling times less than $10^{10}$ years.  If there is
significant internal extinction the plotted relation will underestimate the
stellar mass doubling time for galaxies.}

\figurenum{22}
\figcaption[cowie.fig22.ps]
{\label{fig:22}Doubling time versus rest absolute $K$\ magnitude as a function
of redshift for all identified galaxies in the sample with $K \le 20$. The
solid vertical line shows the absolute magnitude detection limit at the center
of the redshift interval, and the dashed horizontal line marks a fiducial
doubling time of $10^{10}$ years, below which value galaxies may be thought of
as ``in formation''.  The various tracks show the evolution of galaxies with
exponentially declining star formation rates over a period of 13~Gyr computed
from the models of Bruzual \& Charlot\protect{\markcite{bru93}} (1993).  
Dotted: $\tau = 3~{\rm Gyr}$, solid: $\tau = 7~{\rm Gyr}$, dashed:  constant
star formation rate.  The migration of rapidly star-forming galaxies from their
original locus across this plane with changing redshift interval suggests that
in this simple model (ignoring internal extinction) the star formation rate is
declining more rapidly than it would for a 3 Gyr burst model.}

\figurenum{23}
\figcaption[cowie.fig23.ps]
{\label{fig:23}$K$-band light density ($K \le 20$\ sample) of galaxies in the
redshift intervals $z = 0 - 0.3$\ (left panel) and $z = 0.3 -1 $\ (right
panel).  Both are computed for $q_0 =0.5$.  The cumulative luminosity function
of Fig.~\protect{\ref{fig:6a}} has been split into samples of red galaxies
($UK(AB) > 1.3$, pluses) with doubling times $t_d > \ten{10}$ years and blue
galaxies ($UK(AB) < 1.3$, filled squares) with doubling times $t_d < \ten{10}$
years.  The fraction of light in forming objects is considerably higher for
the higher redshift interval (right panel), and extends to brighter galaxy
magnitudes.  The solid curve shows the shape of the cumulative luminosity
density from the forming galaxies of the lower redshift interval (left panel)
renormalized to the higher redshift curve, and may be compared with the curve
for forming galaxies (filled squares) in the right panel.}

\figurenum{24}
\figcaption[cowie.fig24.ps]
{\label{fig:24}Rest frame \itk-band light density of blue galaxies as a
function of redshift interval for a $K \le 20.5$ sample.  The smooth increase
in the luminosity of the most luminous forming galaxies with increasing
redshift is readily apparent.}

\figurenum{25}
\figcaption[cowie.fig25.ps]
{\label{fig:25}$B$\ magnitude versus redshift for all $B \leq 24.5$ galaxies
(pluses) in the present sample or in Songaila et al.\protect{\markcite{son94}}
(1994) compared to ``forming'' galaxies (filled squares) chosen either on the
basis of rest-frame UV--IR color (left panel) or rest-frame [O II] equivalent
width (right panel).  The forming galaxies first enter at $B\sim22$, $z\sim0.3$
and then spread to a wide range of redshifts at fainter magnitudes.  This
coincidence is responsible for the simultaneous rise in the $B$\ counts at $B
\gtrsim 23$ and the rapid blueing which sets in at faint $B$\ and
$K$\ magnitudes.}

\figurenum{26}
\figcaption[cowie.fig26.ps]
{\label{fig:26}Galaxy contributions to the ultraviolet sky surface brightness
as a function of redshift, rest-frame galaxy color, and absolute rest
$U$\ magnitude.  The dotted lines show the range of values expected from the
present-day metallicity.  The left two panels show the contributions to the
extragalactic background light (EBL) from forming galaxies ($t_d < \ten{10}$
years) over two intervals for absolute rest $U$\ magnitude.  A comparison of
the solid curves for the left two panels shows again that large galaxies form
earlier than small ones.  The dashed line in the upper left panel shows the
effect of correcting for one magnitude of internal extinction on the
contribution of observed galaxies to the EBL. It is apparent that we are
already seeing a large fraction of the light required to make all observed
galaxies.  The right two panels compare the contributions from all galaxies
and ``forming'' galaxies to the EBL, assuming no extinction.  In the upper
right panel the solid line shows the contribution from all galaxies compared
to that of the `forming' galaxies (dashed line), which are also shown in the
lower right panel.  Again, these curves show that we are observing the bulk
of galaxy formation, with roughly half of the star formation occurring in
rapidly forming galaxies (upper right panel).}

% Dummy tables for cross-referencing
\begin{table}
\tablenum{1a}
\dummytable\label{tbl-1a}
\tablenum{1b}
\dummytable\label{tbl-1b}
\end{table}
 
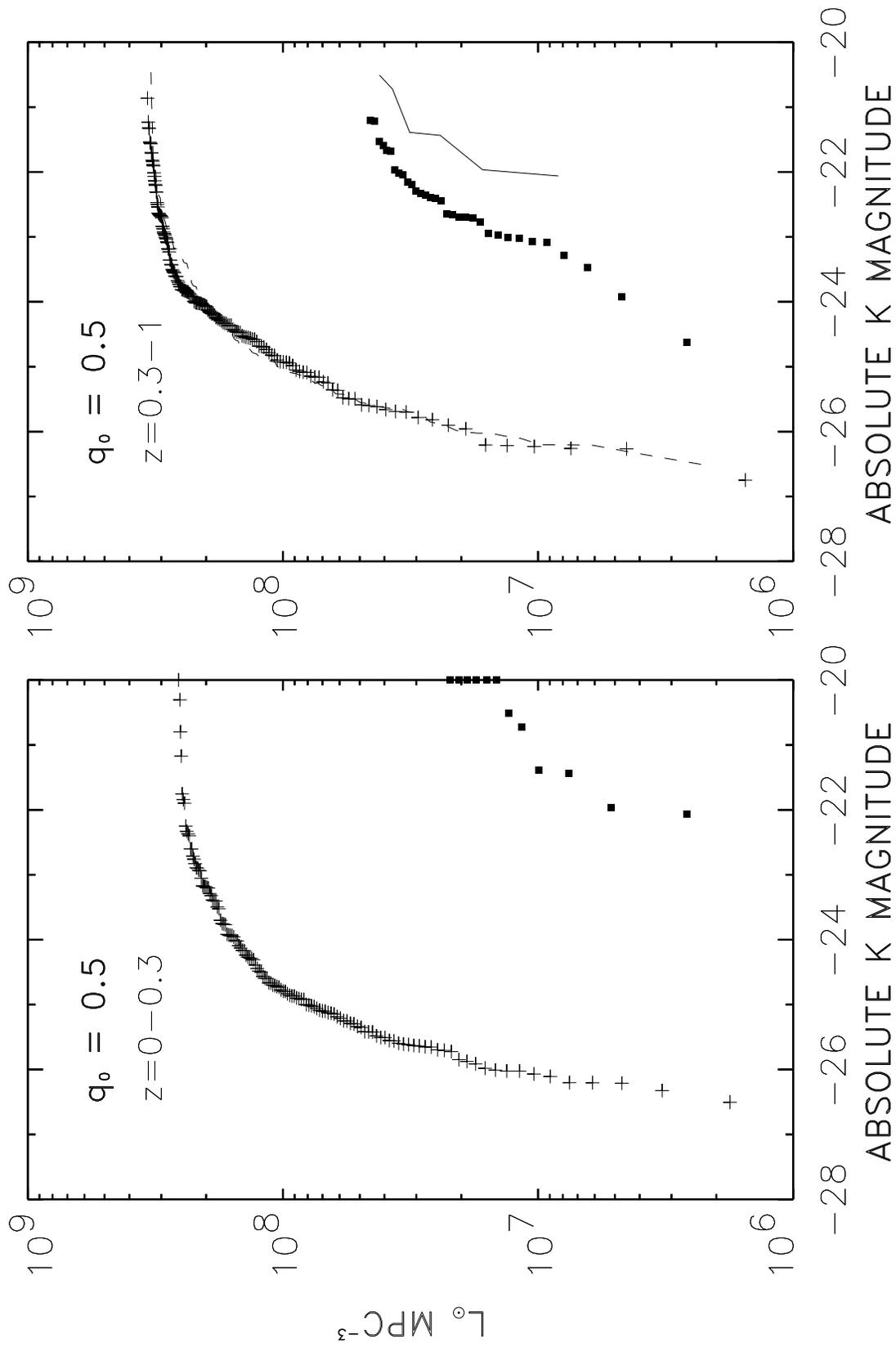
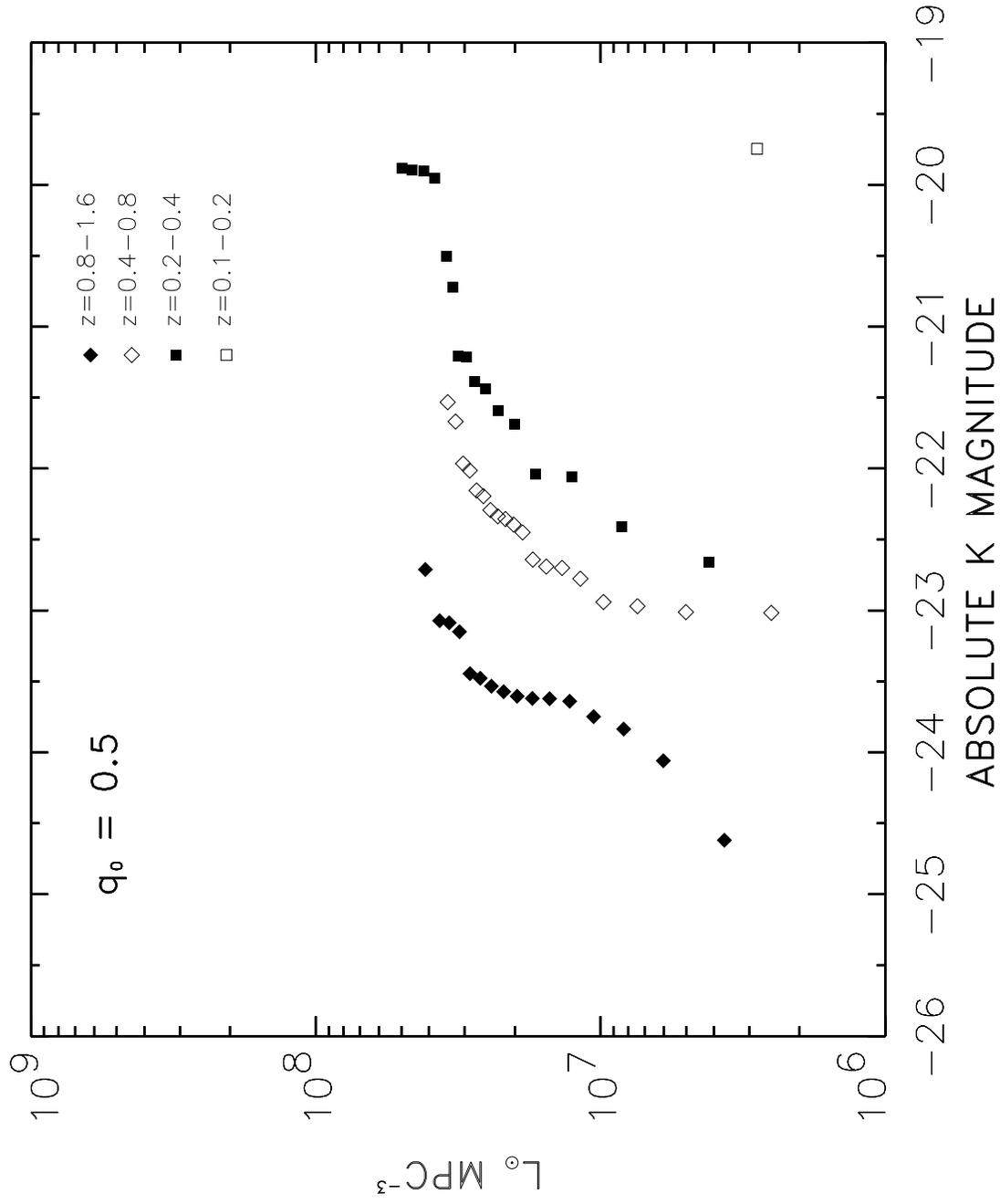
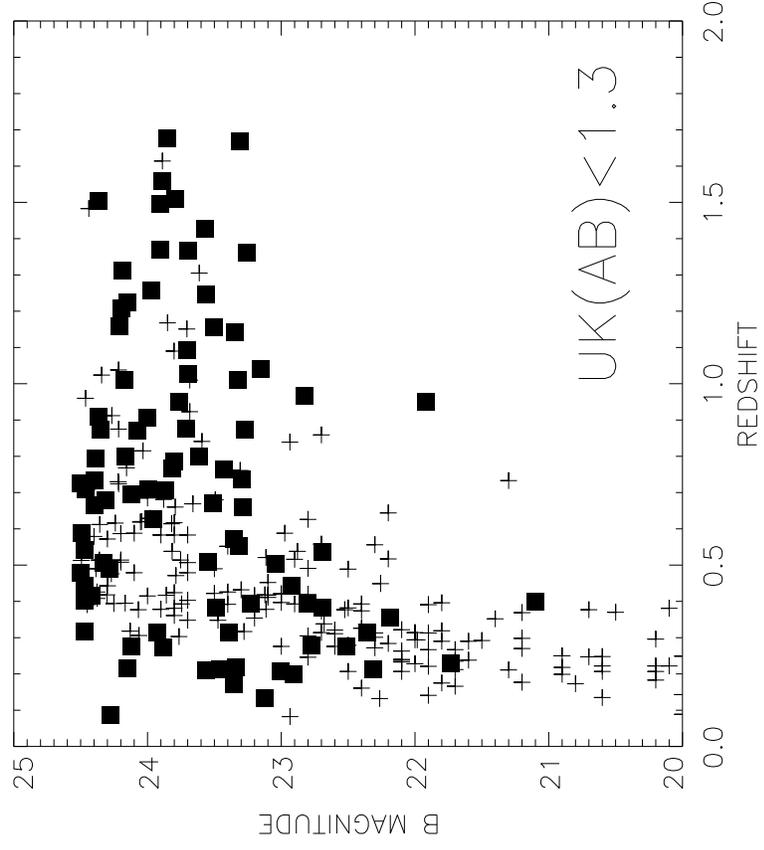
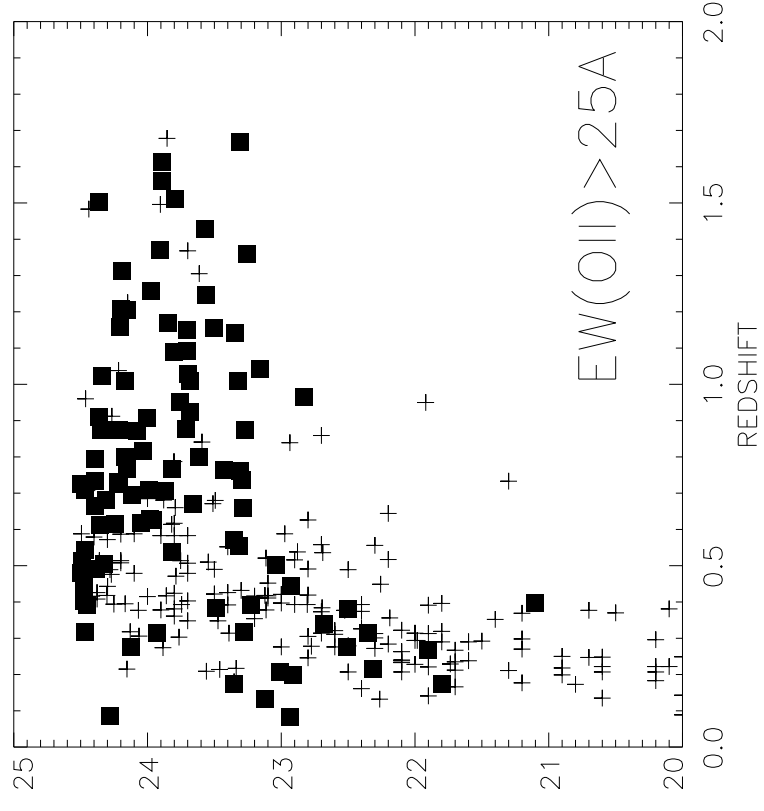
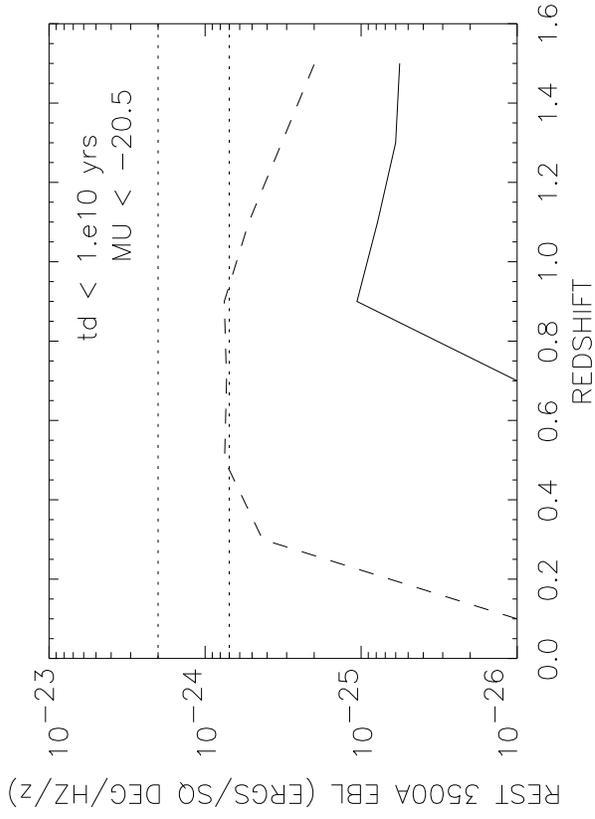
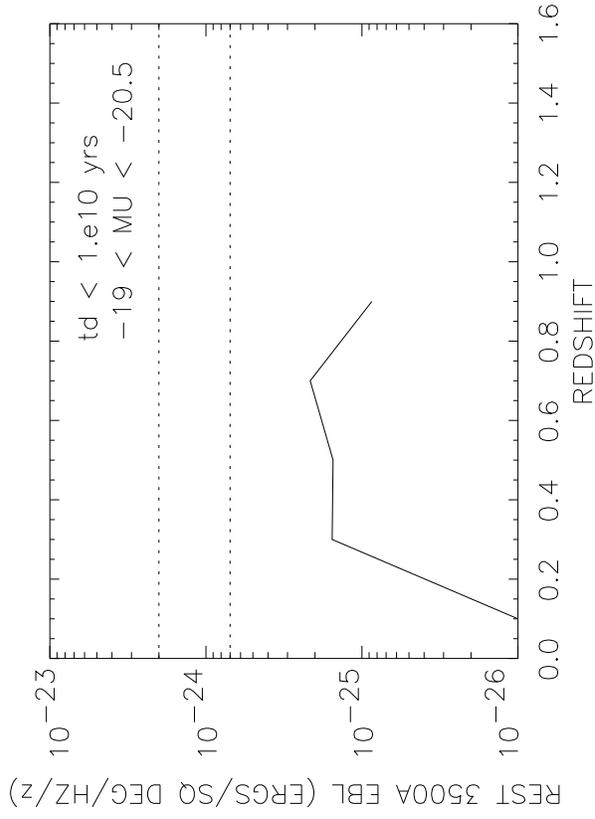
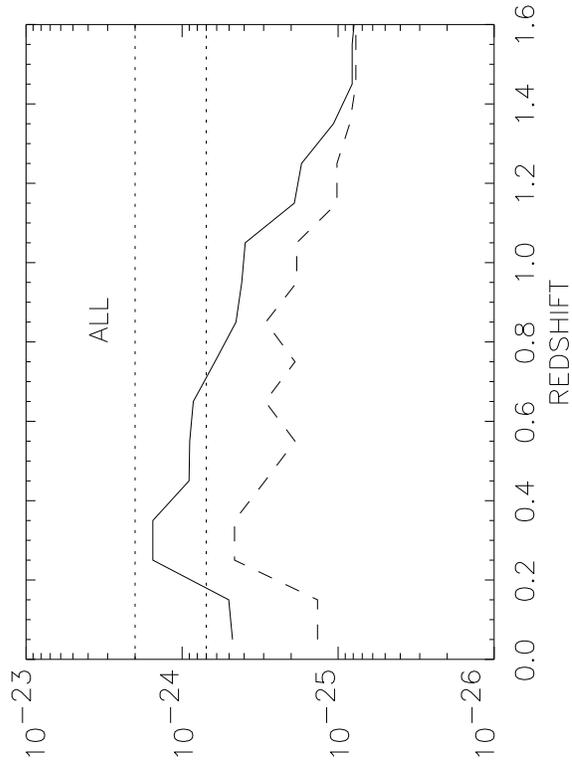
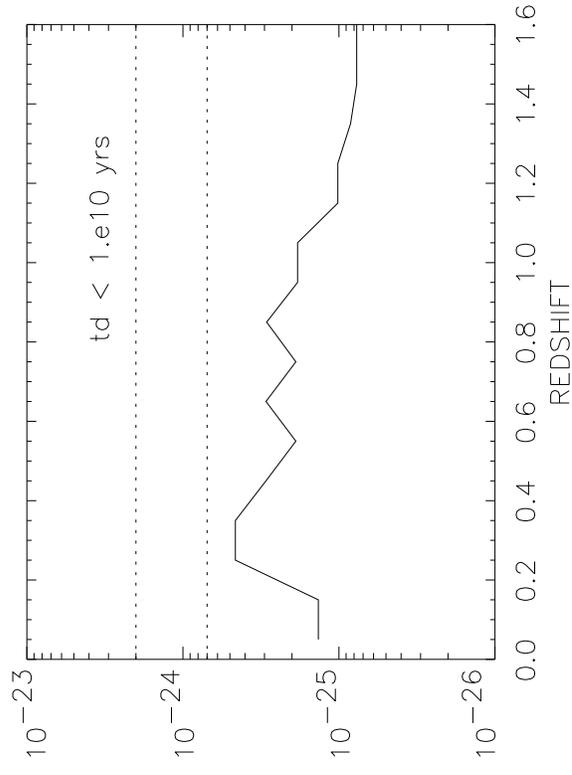
\end{document}